# Bistable optical system based on hysteresis in the reflectivity of graphene-on-Pb(Zr$_x$Ti$_{1-x}$)O$_3$

Strikha M. V.

V. Lashkariov Institute of Semiconductor Physics, National Academy of Sciences of Ukraine, 01650 Kyiv, Ukraine, maksym_strikha@hotmail.com

**Abstract.** We analyse a model describing hysteretic behaviour of the reflectivity $R$ for the system 'graphene–Pb(Zr$_x$Ti$_{1-x}$)O$_3$ (PZT) ferroelectric substrate–gate' with a gate voltage variation, which takes into account trapping of electrons into the graphene–PZT interface states. We demonstrate that the hysteresis in the $R$ parameter can be observed experimentally for the telecommunication-range radiation (the wavelength $\lambda = 1.55$ μm) at low gate voltages and, moreover, the phenomenon can be used while creating fast bistable systems for the novel non-volatile memory devices with on-chip optical interconnection.

Graphene on Pb(Zr$_x$Ti$_{1-x}$)O$_3$ (PZT) has been extensively studied during the recent two years (see, e.g., the review [1] and references therein). PZT as a substrate for the graphene is unique since it opens up the possibilities for both non-volatile gating and extremely high dielectric permittivity (up to 3850 [2] near the morphotropic phase boundary). This allows for reaching the charge densities more than two orders of magnitude higher than those typical for the graphene placed on SiO$_2$ substrates. Up to now robust bistable operation associated with single- and multilayer graphene ferroelectric memory has been reported in the works [3–5]. Some possibilities for creating efficient modulators based on graphene-on-PZT structures for the near-IR and middle-IR radiation and aimed at on-chip optical interconnection applications have been recently discussed in the study [6].

An unusual resistance hysteresis in graphene field-effect transistors fabricated on the ferroelectric PZT has been observed experimentally in the works [3–5]. Fig. 1 presents the sheet resistance $\rho$ as a function of the gate voltage $V_g$ for the seven-layer graphene device, which has been obtained in the study [3] at 300 K. The resistance exhibits distinct behaviours at both low and high $V_g$. When the gate sweep is limited to $|V_g| < 2$ V (see the curve corresponding to the left maximum), the carrier density $n$ and the resistivity follow a conventional field-effect modulation, and the forward and backward sweeps reproduce one another. At $|V_g| > 2$ V, the resistance becomes hysteretic, with the backward sweeping curve maximum being shifted to the right of the forward sweeping curve. A similar hysteresis has been observed in Refs. [3–5], regardless of a specific number of graphene layers ($N = 1$–15), the carrier mobility (16000–140000 cm$^2$/Vs), and the dielectric constant of the PZT (30–500).

There are two characteristic features of the behaviour shown in Fig. 1. First, the direction of the hysteresis is opposite to that expected from the considerations for the charge carrier density induced by polarisation reversal in the PZT. Therefore this behaviour has been referred to as an 'anti-hysteresis' in Ref. [3]. Second, the hysteresis occurs at the $V_g$ value smaller than the coercive voltage necessary to reverse the polarisation in the PZT.

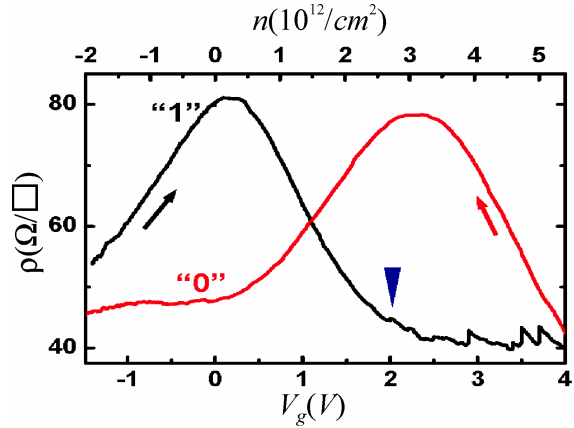

**Fig. 1.** Anti-hysteresis observed in the resistance of graphene-on-PZT [3].

The anti-hysteresis observed in the $\rho$ parameter is reproducible and characterised by long relaxation times. It has been explained in Refs. [3, 5] by screening of the electric field occurring in the PZT substrate by electrons trapped by some states on the grapheme–PZT interface. The numerical model of the process has been suggested in Ref. [7]. The gated single-layer graphene may be examined involving the Fermi energy $E_F$ depending on the concentration according to

$$E_F = \hbar v_F (\pi n)^{1/2}, \tag{1}$$

where $\hbar$ is the Planck constant and $v_F = 10^8$ cm/s. It has been supposed that an interface state exists that has the energy $E_T$. On the forward $V_g$ sweep (when $E_F < E_T$), the carrier concentration $n$ is governed by a simple relation

$$n = \kappa V_g / 4\pi e d, \tag{2}$$

where $d$ denotes the substrate thickness, $\kappa$ the dielectric constant, and $e$ the electron charge. However, when $E_F = E_T$, the electrons from the gated graphene are captured by the interface states of some high 2D density $n_T$. The negative charge of the occupied interface states screens the field in the substrate, so that for the further forward $V_g$ sweep the concentration of carriers in the gated graphene is given by

$$n = \kappa V_g / 4\pi e d - n_T. \tag{3}$$

The next assumption should be that the lifetime of electrons on the interface states is much greater than the switching time of our system. Therefore the relation given by Eq. (3) is also valid for the backward sweep, and the general dependence of $n$ on the gate voltage has a hysteretic shape presented in Fig. 2 (see curves 1 and 2, where the arrows indicate the sweep directions). Now the curve 2 would reach the Dirac point at some gate voltage $V_{DP}$ determined by the concentration $n_T$ of the interface states:

$$V_{DP} = 4\pi e d n_T / \kappa. \tag{4}$$

Notice that, left to the Dirac points, Fig. 2 (curves 1 and 2) presents the holes concentrations. At large negative $V_g$ values, the trapped electrons recombine with the holes in the graphene sheet, so that the $n$ parameter is again governed by Eq. (2).

The total resistivity of the graphene sheet is inversely proportional to its conductivity:
$$\rho(V_g) = 1/(\sigma(V_g) + \sigma_{\min}). \quad (5)$$

Here the first term in denominator represents the gated graphene conductivity, which varies linearly with $V_g$ and $n$, and the second one ($\sigma_{\min} \approx 4e^2/\hbar$) is the minimal graphene's conductivity in the Dirac point [8]. The dependence of $\rho(V_g)$ is also presented in Fig. 2 (here curves 3 and 4 correspond respectively to the forward and backward sweeps). One can see that it has the hysteretic shape observed experimentally in Refs. [3–5]. The distance between the Dirac points in the curves 3 and 4 is determined by the concentration of the interface states via Eq. (4) and, in this rough approximation, it does not depend upon $E_T$. The relaxation of the anti-hysteresis observed experimentally can be explained by a final lifetime of electrons on these states. Of course, a real physical nature of these states remains unclear and needs special examination.

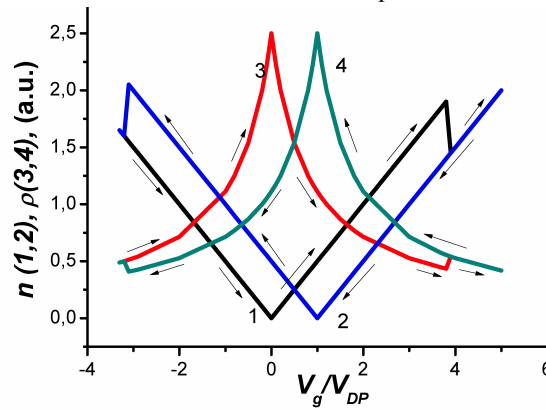

**Fig. 2.** Anti-hysteresis observed for the concentration of carriers (curves 1 and 2) and the resistance (curves 3 and 4) in the graphene-on-PZT: theory. Arrows indicate directions of gate voltage sweeps.

A simple model presented above explains the anti-hysteretic behaviour of the resistivity of graphene-on-PZT. The assumptions made neglect the hysteresis taking place for the PZT itself. However, this can be justified only for the region of small $V_g$, which corresponds to the $n$ values much smaller than the nominal 2D charge density relevant to the polarisation of the PZT ($\sim 3 \times 10^{14}$ cm$^{-2}$ [3]).

A non-volatile graphene-on-ferroelectric memory device has been worked out in the woks [3–5], using a difference between the two resistivity values of the ferroelectric field-effect transistor (see Fig. 1). Below we will demonstrate that the hysteresis in the dependence of the carrier concentration on the gate voltage (Fig. 2) can also be detected using optical methods.

An essential feature of optical properties of the graphene is its substantial interaction with radiation in the wide spectral range (from the far-IR up to the UV one), due to effective interband transitions (see [9–11] and refs therein). A graphene-based optical modulator for the near-IR region (1.35–1.6 μm) has been developed in the work [12]. It has been demonstrated that the modulator can prove promising for the devices with on-chip optical interconnections.

The general theory of the carrier-induced modulation of radiation by a gated graphene has been worked out in Ref. [13]. It has been shown there that the contribution of carriers modifies essentially the response of graphene due to Pauli blocking effect, provided that the absorption is suppressed at $\hbar\omega/2 < E_F$. At low temperatures (or at high doping levels), the threshold frequency for the absorption jump (when the absorption becomes essential) is defined by the condition

$$\hbar\omega_{th} = 2E_F \sim \sqrt{n} .\qquad(6)$$

The transmission and the reflection coefficients of the system 'graphene layer–substrate–Si gate' have been calculated in Ref. [13] for high-$\kappa$ dielectric substrates and in Ref. [6] for the PZT substrate (in all the cases, the geometry of normal propagation of radiation has been examined). The results [6] demonstrate possibilities for fabricating low-voltage gated graphene on the PZT ferroelectric substrates for the modulators for near- and mid-IR regions. When compared with the modulator built in Ref. [12] basing on atomically deposited 7 nm-thick $Al_2O_3$ substrates, the advantage of such a modulator can be comparative simplicity of preparation of epitaxial PZT-film substrates. Another advantage is a possibility for using the reflected wave, which is also intensity-modulated. This can simplify the geometry used in Ref. [12] where the gate has also served as a waveguide.

Now we examine the reflection of the system 'graphene layer–PZT substrate–Si gate' under the condition of normal propagation of light (the wavelength $\lambda$). Let us note that the threshold wavelength $\lambda_{th}$ corresponding to the threshold frequency given by Eq. (6), for which the modulation effect begins, may be written on the basis of Eq. (1) as

$$\lambda_{th} = \frac{\sqrt{\pi}c}{v_F\sqrt{n}} ,\qquad(7)$$

where $c$ is the free-space light velocity. When $n = n_T$, the modulation edge in the $V_g$ scale corresponds to the point $V_{DP}$. Then the reflection coefficient $R$ at this point increases rapidly up to the value approximately equal to 0.023 $N$, where $N$ is the number of graphene layers. With further increase in $V_g$, when $E_F$ becomes equal to $E_T$, the electrons are trapped on the interface levels and the concentration of electrons in the graphene is governed by Eq. (3). Then the $E_F$ parameter decreases and can become smaller than the value corresponding to $V_{DP}$ (see Eqs. 1 and 2). This means that the graphene layer can no longer modulate the radiation with the wavelength $\lambda$, because the direct interband transitions are now forbidden, due to the Pauli blocking effect. Therefore the reflectivity $R$ decreases to its initial level (see Fig. 3, curve 3).

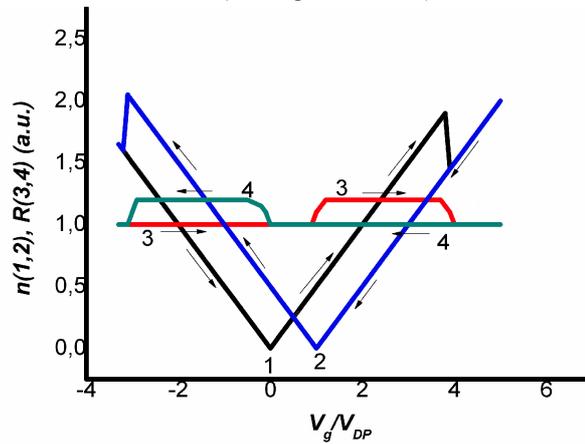

**Fig. 3.** Anti-hysteresis observed for the concentration of carriers (curves 1 and 2) and the reflectivity (curves 3 and 4) in the graphene-on-PZT: theory. Arrows indicate directions of gate voltage sweeps.

For the backward sweep of $V_g$, the concentration $n$ follows the curve 2 in Fig. 3, while the reflectivity $R$ is schematically displayed by the curve 4 of the same figure. This implies that, owing to the interband transitions on holes, the modulation of the radiation with the wavelength $\lambda$ starts at $V_g = 0$ (each of the curves 1 and 2 in Fig. 3 is symmetrical with respect to the corresponding Dirac point (see Eqs. (1)–(4)). The $R$ parameter increases at this point and then decreases back to its initial level when the electrons on the interface states recombine with the holes in the graphene sheet and so the hole concentration decreases.

The hysteresis in the $R$ parameter can be observed experimentally and used for creating fast bistable systems for novel non-volatile memory devices with on-chip optical interconnection. Nonetheless, an important practical problem of applications of the effect is how to modulate the radiation of the telecommunication region ($\lambda = 1.55$ μm) widely used in case of $SiO_2$-based fibres. Substitution of this wavelength value into Eq. (7) and taking into account Eqs. (3) and (4) yield in $n_T \approx 1.2 \times 10^{13}$ cm$^{-2}$. This value seems to be quite realistic for the graphene–ferroelectric interfaces [7, 14] and, moreover, the concentration of this order of magnitude has been observed experimentally in the studies [3–5]. Notice also that estimations following from Eqs. (1)–(4) and (6) testify that the effect does occur at low gate voltages ($V_g = 2–3$ V), which also makes it attractive from the viewpoint of different possible applications. The results obtained should promote further development in the construction of non-volatile graphene-based memory devices of a novel generation.

This work has been supported by the State Fundamental Research Fund of Ukraine under the Grant 40.2/069.

Strikha M. V. , 2012. Bistable optical system based on hysteresis in the reflectivity of graphene-on-Pb(Zr$_x$Ti$_{1-x}$)O$_3$. Ukr.J.Phys.Opt. **13**: 45 –50.

***Анотація.*** *Проаналізовано модель гістерезисної поведінки коефіцієнта відбивання системи „графен–сегнетоелектрична підкладка Pb(Zr$_x$Ti$_{1-x}$)O$_3$ (PZT)–затвор" зі змінною напругою на затворі з урахуванням захоплення електронів на інтерфейсні стани на межі графен–сегнетоелектрик. Показано, що такий гістерезис можна спостерігати експериментально для випромінювання телекомунікаційного діапазону (λ = 1,55 мкм) для низьких напруг на затворі та можна використати у створенні швидкодійної бістабільної системи для нових пристроїв енергонезалежної пам'яті з оптичними з'єднаннями на чіпах.*